\documentclass[notitlepage,a4paper,aps,prd,onecolumn,superscriptaddress,nofootinbib,groupedaddress,longbibliography]{revtex4-1}
\usepackage{amsmath}
\usepackage{amsfonts}
\usepackage{amssymb}
\usepackage{accents}
\usepackage[utf8]{inputenc}
\usepackage[T1]{fontenc}
\usepackage[colorlinks=true]{hyperref}
\usepackage{tikz}
\usetikzlibrary{arrows.meta}

\newcommand{\order}[2]{\accentset{#2}{#1}}
\newcommand{\lc}[1]{\accentset{\circ}{#1}}
\newcommand{\dd}{\mathrm{d}}

\begin{document}

\title{Post-Newtonian limit of generalized symmetric teleparallel gravity}

\author{Kai Flathmann}
\email{kai.flathmann@uni-oldenburg.de}
\affiliation{Institut f\"ur Physik, Universit\"at Oldenburg, 26111 Oldenburg, Germany}

\author{Manuel Hohmann}
\email{manuel.hohmann@ut.ee}
\affiliation{Laboratory of Theoretical Physics, Institute of Physics, University of Tartu, W. Ostwaldi 1, 50411 Tartu, Estonia}

\begin{abstract}
In this article we analyze the post-Newtonian approximation of a generalization of the symmetric teleparallel gravity with the help of the parameterized post-Newtonian (PPN) formalism. This class of theories is based on a free function of the five independent quadratic contractions of the non-metricity tensor. By calculating the PPN metric of these theories, we can restrict the Taylor coefficients of the free function with the help of the PPN parameters and their observational bounds. We find two families of theories whose PPN parameters are identical to those of general relativity, and thus in full agreement with observations. For three further families, we find that only the PPN parameters \(\beta\) and \(\gamma\) deviate, but can be brought arbitrarily close to their general relativity values by an appropriate choice of the Lagrangian, so that also these families contain candidate theories which agree with observations. The remaining theories either possess no well-defined solution of the post-Newtonian field equations, or possess a post-Newtonian limit which exceeds the form assumed in the PPN formalism.
\end{abstract}

\maketitle


\section{Introduction}\label{sec:intro}
During the last century the general theory of relativity, which conventionally attributes the gravitational interaction to the curvature of spacetime, encoded in the presence of the Ricci scalar \(R\) in the action, has passed most of the tests given by solar system and astrophysical observations~\cite{Bertotti:2003rm,Fienga:2011qh,Fienga:2014bvy,Linder:2010py,Olmo:2005zr,Will:2014kxa,Will:2018bme,Will:2018mcj}. Several observations in cosmology and the dynamics of large scale structures, however, are unexplained within general relativity itself, unless it is supplemented with additional, ``dark'' components~\cite{Riess:1998cb,Perlmutter:1998np,Scolnic:2017caz,Aghanim:2019ame,Aghanim:2018eyx,Aghanim:2018oex,Akrami:2019izv,Akrami:2019bkn,Akrami:2018odb,Ahumada:2019vht,Alam:2020sor}. These open questions, as well as the quest for a quantum theory of gravity, have motivated the construction and investigation of several modifications and alternatives to general relativity~\cite{Nojiri:2006ri,Nojiri:2010wj,Nojiri:2017ncd}.

While the aforementioned modifications retain the interpretation of gravity as a manifestation of curvature, this underlying geometrical interpretation of gravity is not without alternatives. Recent developments shed light on two different interpretations of the nature of gravity~\cite{BeltranJimenez:2019tjy}. One alternative is teleparallel gravity, where gravity is a manifestation of a connection on spacetime with vanishing curvature and non-metricity, but non-vanishing torsion~\cite{Aldrovandi:2013wha}. The third possibility, next to curvature and torsion, is to attribute gravity to a flat, torsion-free connection, thus making non-metricity the only non-vanishing tensorial quantity~\cite{Nester:1998mp,Adak:2005cd,Adak:2008gd,Mol:2014ooa,Adak:2018vzk,Jimenez:2019yyx,DAmbrosio:2020nqu}. Most remarkably, the Einstein-Hilbert action of general relativity can be formulated equivalently in either of these other geometries, by introducing suitable scalars \(T\) and \(Q\) from the torsion and non-metricity, respectively, which take the role of the Ricci tensor in the action. This leads to the teleparallel equivalent of general relativity (TEGR)~\cite{Maluf:2013gaa} in the case of torsion, and the symmetric teleparallel equivalent of general relativity (STEGR)~\cite{Nester:1998mp} in the case of non-metricity. Also a more general equivalent, featuring both torsion and non-metricity, has been proposed~\cite{Jimenez:2019ghw}. In these alternative formulations, equivalence is to be understood as dynamical equivalence in the sense that the field equations of all three formulations impose the same dynamics for the metric, which is present in all three geometries.

While general relativity can equivalently be formulated in terms of each of the three quantities - curvature, torsion or non-metricity - generalizations of these three approaches towards gravity are not equivalent anymore. For example, the $F(R)$ class of gravity theories~\cite{Sotiriou:2006hs,DeFelice:2010aj} is not equivalent to their corresponding $F(T)$ gravity theories~\cite{Bengochea:2008gz,Krssak:2015oua,Cai:2015emx}, where $F$ is a free function constituting the Lagrangian, which determines a particular theory within this class. The reason for this is, that the two scalars \(R\) and \(T\) differ by a total derivative, which contributes to the gravitational field equations, unless the Lagrangian function \(F\) is linear, in which case it becomes a boundary term and the theories are dynamically equivalent. For a general function \(F\), however, this difference leads to different field equations, hence to inequivalent dynamics. The same holds true for theories which introduce scalar fields, which are non-minimally coupled to the gravitational action~\cite{Fujii:2003pa,Faraoni:2004pi,EspositoFarese:2000ij,Hohmann:2018rwf,Hohmann:2018vle,Hohmann:2018dqh,Hohmann:2018ijr}. Therefore it is worth studying such kind of modifications, which are based on torsion and non-metricity, as they are essentially different from the class of modifications based on curvature.

The symmetric teleparallel formulation of gravity theories, in which non-metricity is employed as the mediator of gravity, is the least investigated so far, and only a few of its generalizations have been discussed. One possibility is to generalize the symmetric teleparallel gravity action by replacing the non-metricity scalar with an arbitrary function \(F(Q)\), following the same line of thought as the aforementioned \(F(R)\) and \(F(T)\) theories~\cite{BeltranJimenez:2017tkd}. This class of theories has been studied in particular in the context of cosmology~\cite{Jimenez:2019ovq,Lazkoz:2019sjl,Barros:2020bgg,Mandal:2020buf,Mandal:2020lyq}. Another possibility is given by generalizing the non-metricity scalar to other terms quadratic in the non-metricity tensor, known as newer general relativity~\cite{BeltranJimenez:2017tkd,BeltranJimenez:2018vdo,Hohmann:2018wxu,DAmbrosio:2020njs}. Yet another class of theories is obtained by non-minimally coupling scalar fields to non-metricity~\cite{Jarv:2018bgs,Runkla:2018xrv}. Finally, even more general theories have been proposed, such as~\cite{Conroy:2017yln,Koivisto:2018aip,Hohmann:2018xnb,Dialektopoulos:2019mtr,Koivisto:2019jra,Harko:2018gxr,Lobo:2019xwp}.

The large number of gravity theories considered to address the open questions in cosmology requires efficient tools in order to be constrained by observations. One of the most powerful tools to obtain observational bounds on modified theories of gravity is the parametrized post-Newtonian (PPN) formalism \cite{Will:1993ns,Will:2014kxa,Will:2018bme}. It has been widely used to restrict several classes of gravity theories based on curvature~\cite{Hohmann:2013rba,Hohmann:2015kra,Hohmann:2016yfd,Hohmann:2017qje,Nordtvedt:1970uv,Olmo:2005hc,Perivolaropoulos:2009ak,Scharer:2014kya} and torsion~\cite{Emtsova:2019qsl,Flathmann:2019khc,Gladchenko:1990nw,Li:2013oef,Sadjadi:2016kwj,Ualikhanova:2019ygl}, as well as to gravity theories based on multiple metrics~\cite{Clifton:2010hz,Hohmann:2013oca,Hohmann:2017uxe}. Recently, a gauge-invariant approach to the PPN formalism has been developed, which consolidates its mathematical foundation as a tool to study gravity theories by the way how they describe the geometry of spacetime~\cite{Hohmann:2019qgo}.

The purpose of this article is to apply the PPN formalism to a class of symmetric teleparallel gravity theories, thereby opening the path of post-Newtonian calculations for theories in which gravity is modeled by non-metricity. Here we consider a class of theories whose action is determined by a free function of the five parity-even scalars which are quadratic in the non-metricity tensor~\cite{Koivisto:2018aip,Hohmann:2018xnb,Dialektopoulos:2019mtr}. This class of theories is interesting since it encompasses various other theories mentioned above, including newer general relativity and $F(Q)$ gravity~\cite{BeltranJimenez:2017tkd}.

The article is structured as follows. In Sec.~\ref{sec:dynamics} we review the class of symmetric teleparallel gravity theories analyzed in this article. We state its geometric foundations and present the field equations emerging from an action which consists of a free function of the five independent contractions of the non-metricity tensor. Then we state the main assumptions we make in order to be able to apply the PPN formalism in Sec.~\ref{sec:ppn}. In Sec.~\ref{sec:solution} we calculate the post-Newtonian approximation of the field equations and solve them order by order up to the fourth velocity order, where we will be assuming the standard PPN gauge. With the help of these perturbative solutions we calculate the PPN parameters and fully classify the studied theories by the Taylor coefficients of the free function of the Lagrangian in Sec.~\ref{sec:classification}. We summarize our results and give an outlook in Sec.~\ref{sec:conclusion}.

In this article we use Greek letters $\alpha, \beta, \ldots = 0, \ldots, 3$ for spacetime indices and Latin letters $a, b, \ldots = 1, \ldots, 3$ for spatial indices. The sign convention for the flat Minkowski metric is chosen as $(-,+,+,+)$.

\section{Field variables and their dynamics}\label{sec:dynamics}
In this section we review the action of the theory we aim to analyze in this article. The dynamical fields of the theory are given by a Lorentzian metric $g_{\mu\nu}$ and an affine connection $\Gamma^{\rho}{}_{\mu\nu}$. The latter is further restricted by the conditions that it has vanishing curvature
\begin{equation}\label{eqn:nocurv}
R^{\rho}{}_{\sigma\mu\nu} = \partial_{\mu}\Gamma^{\rho}{}_{\sigma\nu} - \partial_{\nu}\Gamma^{\rho}{}_{\sigma\mu} + \Gamma^{\rho}{}_{\lambda\mu}\Gamma^{\lambda}{}_{\sigma\nu} - \Gamma^{\rho}{}_{\lambda\nu}\Gamma^{\lambda}{}_{\sigma\mu} = 0\,,
\end{equation}
and vanishing torsion
\begin{equation}\label{eqn:notors}
T^{\rho}{}_{\mu\nu} = \Gamma^{\rho}{}_{\nu\mu} - \Gamma^{\rho}{}_{\mu\nu} = 0\,.
\end{equation}
Nevertheless, we allow for non-vanishing non-metricity
\begin{equation}
Q_{\rho\mu\nu}=\nabla_{\rho}g_{\mu\nu}\,.
\end{equation}
Combining the first two conditions, it follows the affine connection has to be of the form
\begin{equation}\label{eqn:flatsymconn}
\Gamma^{\rho}{}_{\mu\nu}=(\Lambda^{-1})^{\rho}{}_{\lambda}\partial_{\nu}\Lambda^{\lambda}{}_{\mu}\,,
\end{equation}
with $\partial_{[\mu}\Lambda^{\lambda}{}_{\nu]}=0$. Later, we will use this form of the affine connection for its post-Newtonian expansion. We assume an action of the form
\begin{equation}
S[g,\Gamma,\chi]=S_g[g,\Gamma]+S_m[g,\chi]\,,
\end{equation}
with the gravitational part of the action $S_g$ being of the form
\begin{equation}
S_g[g\,,\Gamma]=\int_{M}\mathcal{F}\left(\mathcal{Q}_1,\mathcal{Q}_2,\mathcal{Q}_3,\mathcal{Q}_4,\mathcal{Q}_5\right)\sqrt{-g}\,\dd^4x\,.
\end{equation}
Here the Lagrangian is given by a free function $\mathcal{F}$ of the five non-trivial quadratic contractions of the non-metricity tensor $Q_{\rho\mu\nu}$
\begin{equation}
\mathcal{Q}_1=Q^{\rho\mu\nu}Q_{\rho\mu\nu}\,, \qquad \mathcal{Q}_2=Q^{\mu\nu\rho}Q_{\rho\mu\nu}\,, \qquad
\mathcal{Q}_3=Q^{\rho\mu}{}_{\mu}Q_{\rho\nu}{}^{\nu}\,, \qquad
\mathcal{Q}_4=Q^{\mu}{}_{\mu\rho}Q_{\nu}{}^{\nu\rho}\,,
\qquad
\mathcal{Q}_5=Q^{\mu}{}_{\mu\rho}Q^{\rho\nu}{}_{\nu}\,.
\end{equation}
The matter action $S_m$ depends on the metric and an arbitrary set of matter fields $\chi$. With that knowledge we can obtain the usual energy-momentum tensor $\Theta^{\mu\nu}$ by varying the matter action $S_m$ with respect to the metric $g_{\mu\nu}$ as
\begin{equation}
\delta S_m[g, \chi] = -\frac{1}{2}\int_{M}\Theta^{\mu\nu}\delta g_{\mu\nu}\sqrt{-g}\,\dd^4 x\,,
\end{equation}
where $g$ denotes the metric determinant.
Throughout this article we will assume for the matter source a perfect fluid, as usual. If we define the derivatives of the free function $\mathcal{F}$ with respect to the five scalar quantities $\mathcal{Q}_i$ as
\begin{equation}
\mathcal{F}_{,i}=\frac{\partial \mathcal{F}}{\partial\mathcal{Q}_i}\,, \qquad i=1,\ldots, 5\,,
\end{equation}
then we can write the field equations as $E_{\mu\nu}=0$, with
\begin{align}\label{eqn:fieldequation}
E_{\mu\nu}= &-2\lc{\nabla}_{\rho}\left(\mathcal{F}_{,1}Q^{\rho}{}_{\mu\nu}+\mathcal{F}_{,2}Q_{(\mu\nu)}{}_{\rho}+\mathcal{F}_{,3}Q^{\rho\sigma}{}_{\sigma}g_{\mu\nu}+\mathcal{F}_{,4}Q^{\sigma}{}_{\sigma(\mu}\delta^{\rho}_{\nu)}\right)\nonumber\\
&-\lc{\nabla}_{\rho}\left[\mathcal{F}_{,1}\left(Q_{\sigma}{}^{\sigma\rho}g_{\mu\nu}+\delta^{\rho}_{(\mu}Q_{\nu\sigma)}{}^{\sigma}\right)\right]+\frac{1}{2}\mathcal{F}g_{\mu\nu}-\mathcal{F}_{,3}Q_{\mu\rho}{}^{\rho}Q_{\nu\sigma}{}^{\sigma}\nonumber\\
&+\mathcal{F}_{,2}\left(2Q^{\rho\sigma}{}_{\mu}Q_{\sigma\rho\nu}-Q_{\mu}{}^{\rho\sigma}Q_{\nu\rho\sigma}-2Q^{\rho\sigma}{}_{(\mu}Q_{\nu)\rho\sigma}\right)\nonumber\\
&+\mathcal{F}_{,4}\left[Q_{\rho}{}^{\rho\sigma}\left(Q_{\sigma\mu\nu}-2Q_{(\mu\nu)\sigma}\right)+Q^{\rho}{}_{\rho\mu}Q^{\sigma}{}_{\sigma\nu}-Q^{\rho}{}_{\rho(\mu}Q_{\nu)\sigma}{}^{\sigma}\right]\nonumber\\
&+\frac{1}{2}\mathcal{F}_{,5}\left[Q^{\rho\sigma}{}_{\sigma}\left(Q_{\rho\mu\nu}-2Q_{(\mu\nu)\sigma}\right)-Q_{\mu\rho}{}_{\rho}Q_{\nu\sigma}{}_{\sigma}\right]-\kappa^2\Theta_{\mu\nu}\,.
\end{align}
Note that covariant derivatives $\lc{\nabla}$ with respect to the Levi-Civita connection are denoted by a circle, to distinguish them from the covariant derivative \(\nabla\) of the independent connection \(\Gamma^{\mu}{}_{\nu\rho}\).

\section{Post-Newtonian approximation}\label{sec:ppn}
In this article we make use of the parameterized post-Newtonian (PPN) formalism~\cite{Will:1993ns,Will:2014kxa,Will:2018bme}. Therefore, we review the general assumptions of the formalism in this section, and supplement them with further assumptions on the independent connection \(\Gamma^{\mu}{}_{\nu\rho}\), in order to apply the PPN formalism to the symmetric teleparallel gravity theory we study here. In the previous section we mentioned, that the matter field will be assumed as a perfect fluid
\begin{equation}\label{eqn:tmunu}
\Theta^{\mu\nu} = (\rho + \rho\Pi + p)u^{\mu}u^{\nu} + pg^{\mu\nu}\,.
\end{equation}
Here we denote the rest energy density, pressure, four-velocity and specific internal energy by $\rho$, $p$, $u^{\mu}$ and $\Pi$, respectively. For the four-velocity we assume the normalization $u^{\mu}u^{\nu}g_{\mu\nu}=-1$. Compared to the speed of light $c \equiv 1$, the velocity $v^i=u^i/u^0$ of the matter is assumed to be small in a given reference frame. As usual we can perturbatively expand the dynamical fields in orders of the velocity $\mathcal{O}(n) \propto |\vec{v}|^n$. We expand the metric around a flat Minkowski background $\eta_{\mu\nu}=\mathrm{diag}(-1, 1, 1, 1)$
\begin{equation}\label{eqn:metricperturb}
g_{\mu\nu}=\eta_{\mu\nu}+h_{\mu\nu}=\eta_{\mu\nu}+\order{h}{2}_{\mu\nu}+\order{h}{3}_{\mu\nu}+\order{h}{4}_{\mu\nu}+\mathcal{O}(5)\,.
\end{equation}
In order to approximate the coefficients \(\Gamma^{\rho}{}_{\mu\nu}\) of the symmetric teleparallel connection, we use the relation~\eqref{eqn:flatsymconn}, which follows from the condition~\eqref{eqn:nocurv} of vanishing curvature. From the condition~\eqref{eqn:notors} of vanishing torsion further follows that the transformation matrices \(\Lambda^{\mu}{}_{\nu}\) are generated by a coordinate transformation,
\begin{equation}\label{eqn:lambdaalphabeta}
\Lambda^{\alpha}{}_{\beta} = \frac{\partial x'^{\alpha}}{\partial x^{\beta}}\,,
\end{equation}
where the coordinates \(x'^{\mu}\) correspond to the so-called coincident gauge~\cite{BeltranJimenez:2017tkd}. This coordinate system is characterized by the property that the connection coefficients \(\Gamma'^{\rho}{}_{\mu\nu}\) vanish. Indeed, from the usual coordinate transformation of connection coefficients one finds the relation
\begin{equation}
\Gamma^{\rho}{}_{\mu\nu} = \Gamma'^{\gamma}{}_{\alpha\beta}\frac{\partial x^{\rho}}{\partial x'^{\gamma}}\frac{\partial x'^{\alpha}}{\partial x^{\mu}}\frac{\partial x'^{\beta}}{\partial x^{\nu}} + \frac{\partial x^{\rho}}{\partial x'^{\gamma}}\frac{\partial x'^{\gamma}}{\partial x^{\mu}\partial x^{\nu}} = \Gamma'^{\gamma}{}_{\alpha\beta}(\Lambda^{-1})^{\rho}{}_{\gamma}\Lambda^{\alpha}{}_{\mu}\Lambda^{\beta}{}_{\nu} + (\Lambda^{-1})^{\rho}{}_{\gamma}\partial_{\nu}\Lambda^{\gamma}{}_{\mu}\,,
\end{equation}
which reduces to the relation~\eqref{eqn:flatsymconn} for \(\Gamma'^{\gamma}{}_{\alpha\beta} = 0\). In the next step, we must assume a background around which we will expand the connection, in analogy to the Minkowski background \(\eta_{\mu\nu}\) in the perturbative expansion~\eqref{eqn:metricperturb} of the metric. Here we assume that this background is simply given by the coincident gauge, hence the connection coefficients vanish. It follows that we can approximate the coordinate transformation in the form
\begin{equation}
x'^{\mu} = x^{\mu} + \xi^{\mu} + \frac{1}{2}\xi^{\nu}\partial_{\nu}\xi^{\mu}
\end{equation}
up to quadratic order in the coefficients \(\xi^{\mu}\), which are the generators of a ``knight diffeomorphism''~\cite{Bruni:1996im,Sonego:1997np,Bruni:1999et}. The coordinate transformation matrices thus take the form
\begin{equation}
\Lambda^{\alpha}{}_{\beta} = \delta^{\alpha}_{\beta} + \partial_{\beta}\xi^{\alpha} + \frac{1}{2}\partial_{\beta}(\xi^{\gamma}\partial_{\gamma}\xi^{\alpha})\,.
\end{equation}
Combining this relation with the form~\eqref{eqn:flatsymconn} of the connection coefficients leads to their expression
\begin{equation}
\Gamma^{\rho}{}_{\mu\nu} = \partial_{\mu}\partial_{\nu}\xi^{\rho} + \frac{1}{2}\left(\xi^{\sigma}\partial_{\mu}\partial_{\nu}\partial_{\sigma}\xi^{\rho} + 2\partial_{(\mu}\xi^{\sigma}\partial_{\nu)}\partial_{\sigma}\xi^{\rho} - \partial_{\mu}\partial_{\nu}\xi^{\sigma}\partial_{\sigma}\xi^{\rho}\right)\,.
\end{equation}
Then we expand $\xi^{\alpha}$ similar to the metric as
\begin{equation}
\xi^{\alpha}=\order{\xi}{2}^{\alpha}+\order{\xi}{3}^{\alpha}+\order{\xi}{4}^{\alpha}+\mathcal{O}\left(5\right)\,.
\end{equation}
The non-vanishing components of the dynamical fields $g$ and $\xi$ we have to calculate are
\begin{equation}
\order{h}{2}_{00}\,,\order{h}{2}_{ij}\,,\order{h}{3}_{i0},\order{h}{4}_{00}\,,\order{\xi}{2}^{i}\,,\order{\xi}{3}^{0}\,,\order{\xi}{4}^{i}\,.
\end{equation}
In order to apply the post-Newtonian approximation to the geometry part of the field equations \eqref{eqn:fieldequation}, we have to expand the free function $\mathcal{F}$ and its derivatives $\mathcal{F}_{,i}$ as a Taylor series
\begin{equation}
\mathcal{F}= F_0+\sum_{k=1}^{5}F_k\mathcal{Q}_k+\mathcal{O}\left(5\right)\,, \quad
\mathcal{F}_{,i}= F_i+ \sum_{k=1}^{5}F_{ik}\mathcal{Q}_{k}+\mathcal{O}\left(5\right)\,,
\end{equation}
where the Taylor coefficients $F_i$ and $F_{ik}$ are assumed to be of velocity order $\mathcal{O}\left(0\right)$ and are calculated at $\mathcal{Q}_i=0$. In the following section we will see that the second order Taylor coefficients \(F_{ik}\) couple to terms that are of higher velocity order than the post-Newtonian approximation and therefore do not contribute to the perturbative equations we consider. For later use, we introduce another parametrization for the linear order Taylor coefficients given by
\begin{subequations}\label{eqn:paramrelat}
\begin{align}
F_1 &= 3C_5\,, & C_1 &= F_2 - F_4\,,\\
F_2 &= \frac{1}{2}(C_1 + C_2 + C_3 - 2C_4 - 4C_5)\,, & C_2 &= \frac{F_1}{3} + F_3\,,\\
F_3 &= C_2 - C_5\,, & C_3 &= F_1 + F_2 + F_3 + F_4 + F_5\,,\\
F_4 &= \frac{1}{2}(-C_1 + C_2 + C_3 - 2C_4 - 4C_5)\,, & C_4 &= F_3 + \frac{F_5}{2}\,,\\
F_5 &= 2(-C_2 + C_4 + C_5)\,, & C_5 &= \frac{F_1}{3}\,.
\end{align}
\end{subequations}
This reparametrization will simplify the field equations and their solutions, and further be helpful for the classification of theories in section~\ref{sec:classification}.

Finally, we can expand the energy-momentum tensor as
\begin{subequations}\label{eqn:energymomentum}
\begin{align}
\Theta_{00} &= \rho\left(1 + \Pi + v^2 - \order{h}{2}_{00}\right) + \mathcal{O}(6)\,,\\
\Theta_{0j} &= -\rho v_j + \mathcal{O}(5)\,,\\
\Theta_{ij} &= \rho v_iv_j + p\delta_{ij} + \mathcal{O}(6)\,.
\end{align}
\end{subequations}
using the standard assumption that the matter variables are of the velocity orders \(\rho \sim \Pi \sim \mathcal{O}(2)\) and \(p \sim \mathcal{O}(4)\). Further, time derivatives are weighted with \(\partial_0 \sim \mathcal{O}(1)\).

\section{Solving the field equations}\label{sec:solution}
Now we can use all expressions calculated in the preceding section to derive the post-Newtonian approximation of the field equations~\eqref{eqn:fieldequation} and their solution. This will be done in several steps. We start with the zeroth velocity order \(\mathcal{O}(0)\), which corresponds to the vacuum equations, in section~\ref{ssec:order0}. Then we proceed with the second velocity order \(\mathcal{O}(2)\) in section~\ref{ssec:order2}, the third velocity order \(\mathcal{O}(3)\) in section~\ref{ssec:order3}, and finally the fourth velocity order \(\mathcal{O}(4)\) in section~\ref{ssec:order4}.

\subsection{Zeroth velocity order}\label{ssec:order0}
The zeroth velocity order of the energy-momentum tensor vanishes identically for all components $\order{\Theta}{0}_{\mu\nu}=0$ and the geometrical part of the field equations~\eqref{eqn:fieldequation} can be calculated at the background $\order{h}{0}_{\mu\nu}=\eta_{\mu\nu}$ and $\order{\xi}{0}^{\alpha}=0$ leading to
\begin{equation}
\order{E}{0}_{\mu\nu}=\frac{1}{2}F_0\eta_{\mu\nu}=0\,.
\end{equation}
Therefore the vacuum field equations can only be solved with our post-Newtonian approximation if $F_0=0$. This result is not surprising, since $F_0$ should be related to a cosmological constant. This would lead to a contradiction to the assumption of an asympotically flat post-Newtonian metric. Hence, we restrict our discussion to this case throughout the remainder of this article.

\subsection{Second velocity order}\label{ssec:order2}
The second velocity order of the field equations~\eqref{eqn:fieldequation} are given by $\order{E}{2}_{00}=0$, $\order{E}{2}_{ij}=0$, with
\begin{align}\label{eqn:fieldeqn_order2}
\order{E}{2}_{00}= \ &\triangle\left[-\left(C_2+2C_5\right)\order{h}{2}_{00}.\left(C_2-C_5\right)\order{h}{2}^a{}_a+2C_4\partial_a\order{\xi}{2}^a\right]-\left(C_2-C_4-C_5\right)\partial_b\partial_a\order{h}{2}^{ab}-\kappa^2\rho\,,\nonumber\\
\order{E}{2}_{ij}=\ &\triangle\left[-\left(C_2-C_5\right)\delta_{ij}\order{h}{2}_{00}-3C_5\order{h}{2}_{ij}-\left(C_2-C_5\right)\delta_{ij}\order{h}{2}^a{}_a+\left(C_2+C_3-2C_4\right)\partial_{(i}\order{\xi}{2}^{j)}+2C_4\delta_{ij}\partial_a\order{\xi}{2}^a\right]\nonumber\\
&+\partial_i\partial_j\left[\left(C_2-C_4\right)\order{h}{2}_{00}+\left(C_2-C_4\right)\order{h}{2}^a{}_a-\left(C_2-C_3\right)\partial_a\order{\xi}{2}^a\right]+\left(C_2-C_5\right)\delta_{ij}\partial_a\partial_b\order{h}{2}^{ab}\nonumber\\
&-\left(C_2+C_3-2C_4\right)\partial_a\partial_{(i}\order{h}{2}^a{}_{j)}
\end{align}
and $\triangle=\partial_a\partial^a$ is the flat space Laplacian. These equations can be solved by introducing the Newtonian potential and the so-called superpotential, which are defined by
\begin{equation}
U=\int\dd^3x'\frac{\rho'}{|\vec{x}-\vec{x}'|}\,, \quad
\chi=-\int\dd^3x'\rho'|\vec{x}-\vec{x}'|\,.
\end{equation}
First we make the ansatz
\begin{equation}\label{eqn:metorder2}
\order{h}{2}_{00} = a_1U\,, \quad
\order{h}{2}_{ij} = a_2\delta_{ij}U\,, \quad
\order{\xi}{2}^i = a_3\partial^i\chi
\end{equation}
to reformulate Eqn.~\eqref{eqn:fieldeqn_order2} as an algebraic equation for the three coefficients $a_i$. By making use of the identity $\triangle\chi=-2U$ and demanding that the coefficients in front of $\partial_i\partial_jU$ and $\delta_{ij}$ vanish, we can derive a linear system of equations written as
\begin{equation}
M^{ij}a_j=0\,,
\end{equation}
where the coefficients of the matrix \(M\) are linear in the parameters \(C_i\). The three coefficients $a_i$ cannot be determined for all $C_i$ unambiguously, but only in the cases in which \(M\) is non-degenerate. To discuss this we make use of the determinant of $M$
\begin{equation}
\det M \propto C_5\left(C_2C_3-C_4^2\right)\,.
\end{equation}
In the case of $C_5=0$, we can only find vacuum (i.e., $\rho=0$) as a solution and $\order{h}{2}$ cannot be determined unambiguously. Hence, we will exclude this case from now on, and assume \(C_5 \neq 0\). If we assume $C_2C_3-C_4^2=0$, $\order{\xi}{2}^i$ cannot be determined uniquely. Keeping this assumption, we can further distinguish the two cases $C_3=0$ and $C_3\neq 0$. In the first case we can determine $\order{h}{2}$ independent from the tensor field $\order{\xi}{2}^i$. In the second case the metric components cannot be determined uniquely. If $\det M\neq 0$ we can solve for all $a_i$. Thus, we have the following two relevant cases which we will further discuss in detail:
\begin{enumerate}
\item
$C_5\neq 0$ and $C_2C_3-C_4^2\neq 0$: We can solve the second order equations for all perturbations and obtain
\begin{equation}
a_1 = \frac{\kappa^2}{72\pi}\frac{2C_4^2 - C_3(2C_2 + C_5)}{C_5\left(C_2C_3 - C_4^2\right)}\,, \quad
a_2 = \frac{\kappa^2}{72\pi}\frac{2C_4^2 - C_3(C_2 - C_5)}{C_5\left(C_2C_3 - C_4^2\right)}\,, \quad
a_3 = \frac{\kappa^2}{288\pi}\frac{C_2C_3 - C_3C_5 - C_4^2 +3C_4C_5}{C_5\left(C_2C_3 - C_4^2\right)}\,.
\end{equation}

\item
$C_5\neq 0$ and $C_3=C_4=0$: The equations degenerate, and we can solve only for the metric potentials, where we find the solution
\begin{equation}
a_1 = -\frac{\kappa^2}{72\pi}\frac{2C_2 + C_5}{C_2C_5}\,, \quad
a_2 = \frac{\kappa^2}{72\pi}\frac{C_5 - C_2}{C_2C_5}\,.
\end{equation}
\end{enumerate}
Note that the result for \(a_1\) and \(a_2\) in the second case is obtained from the first case in the limit \(C_4 \to 0\), in which \(C_3\) cancels.

\subsection{Third velocity order}\label{ssec:order3}
For the third velocity order the only non-vanishing components of the field equations~\eqref{eqn:fieldequation} are $\order{E}{3}_{0i}=\order{E}{3}_{i0}=0$, which are given by
\begin{align}\label{eqn:fieldeqn_order3}
\order{E}{3}_{0i}=\ & \Delta\left[-6C_5\order{h}{3}_{0i}+\left(C_2+C_3-2C_4+2C_5\right)\left(\partial_i\order{\xi}{3}^0+\partial_t\order{\xi}{2}^i\right)\right]\nonumber\\
&+\partial_j\partial_t\left[-2\left(C_2-C_3+2C_5\right)\partial_a\order{\xi}{2}^a+\left(C_2-C_3+2C_5\right)\order{h}{2}_{00}+2\left(C_2-C_4-C_5\right)\order{h}{2}^a{}_a\right]\nonumber\\
&-\left(C_2+C_3-2C_4-4C_5\right)\partial_a\partial_i\order{h}{3}_0{}^a-\left(C_2+C_3-2C_4-4C_5\right)\partial_a\partial_t\order{h}{2}^a{}_i+2\kappa^2\rho v_i\,.
\end{align}
We can solve this equation by introducing the PPN potentials $V_i$ and $W_i$ with
\begin{equation}
V_i=\ \int\dd^3 x'\frac{\rho'v_i'}{|\vec{x}-\vec{x}'|}\,, \quad
W_i=\ \int\dd^3 x'\frac{\rho'v_j'\left(x_i-x'_i\right)\left(x_j-x'_j\right)}{|\vec{x}-\vec{x}'|}\,.
\end{equation}
Now we substitute the ansatz
\begin{equation}\label{eqn:metorder3}
\order{h}{3}_{i0} = \order{h}{3}_{0i} = a_VV_i + a_WW_i\,, \quad
\order{\xi}{3}^0 = a_0\partial^0\chi
\end{equation}
into Eqn.~\eqref{eqn:fieldeqn_order3}. Here $a_V$, $a_W$ and $a_0$ are real constants. The solution is given up to a gauge constant $a_0$, and can most simply be expressed in the linear combination \(a_V + a_W\) and \(a_V - a_W\). The non-degenerate case with $C_5\neq 0$ and $C_2C_3-C_4^2\neq 0$ reads
\begin{equation}\label{eqn:solOrder3_generic}
a_V + a_W = -\frac{\kappa^2}{144\pi}\left(\frac{6}{C_5} + \frac{C_3 + 3C_4}{C_2C_3 - C_4^2}\right)\,, \quad
a_V - a_W = 2a_0\,.
\end{equation}
The case with $C_3=C_4=0$ is tremendously simpler but has an additional ambiguity due to the fact that $\order{\xi}{2}$ cannot be determined, hence leaving the undetermined constant \(a_3\) in the solution
\begin{equation}\label{eqn:solOrder3_degenerate}
a_V + a_W = \frac{\kappa^2}{24\pi C_5}\,, \quad
a_V - a_W = 2(a_0 - a_3)\,.
\end{equation}
The gauge constant $a_0$ in Eqn.~\eqref{eqn:solOrder3_generic}, and correspondingly \(a_0 - a_3\) in Eqn.~\eqref{eqn:solOrder3_degenerate}, will be determined by demanding the standard PPN gauge in the fourth velocity order solution in the following section.

\subsection{Fourth velocity order}\label{ssec:order4}
In order to solve the fourth order equations $\order{E}{4}_{00}=\order{E}{4}_{ij}=0$, which are obtained from the field equations~\eqref{eqn:fieldequation} and which we omit here for brevity, we have to decouple the spatial and time components of the fourth order metric components $\order{h}{4}_{00}$ and $\order{h}{4}_{ij}$. We will perform this decoupling as follows. In order to obtain independent equations and separate the variables, we take the second derivatives of the fourth order equations $\order{E}{4}_{00}=0$, $\order{E}{4}_{ij}=0$, so that we obtain the equations
\begin{equation}
\triangle\order{E}{4}_{00} = 0\,, \quad
\triangle\order{E}{4}^i{}_{i} = 0\,, \quad
\partial_i\partial_j\order{E}{4}^{ij} = 0\,.
\end{equation}
We then have to eliminate the terms $\triangle\triangle\order{h}{4}^i{}_i$, $\triangle\partial_i\partial_j\order{h}{4}^{ij}$ and $\triangle\triangle\partial_i\order{\xi}{4}^i$ from these equations. In the remaining equation, the only unknown we must solve for then appears in the term \(\triangle\triangle\order{h}{4}_{00}\), and it can be solved using the general ansatz
\begin{multline}\label{eqn:generppn}
\triangle\triangle\order{h}{4}_{00} = a_4\triangle p+a_5\triangle(\Pi\rho)+a_{6}\triangle(\rho v_a v^a)+a_{7}\triangle\chi\triangle\triangle\triangle\chi+a_{8}\partial_a\partial_b\chi\partial^a\partial^b\triangle\triangle\chi+a_{9}\partial_a\triangle\chi\partial^a\triangle\triangle\chi\\
+a_{10}\partial_a\partial_b\partial_c\chi\partial^a\partial^b\partial^c\triangle\chi+a_{11}\triangle\triangle\chi\triangle\triangle\chi+a_{12}\partial_a\partial_b\triangle\chi\partial^a\partial^b\triangle\chi+a_{13}\partial_a\partial_b\partial_c\partial_d\chi\partial_a\partial_b\partial_c\partial_d\chi + a_{14}\partial_a\partial_b(\rho v^av^b)\,,
\end{multline}
which depends on constants \(a_4, \ldots, a_{14}\) which are to be determined. Note that the terms entering with $a_4$ to $a_{12}$ and $a_{14}$ can be identified with the standard PPN potentials of the fourth order. The remaining terms may give rise to additional potential which do not occur in the standard PPN formalism, depending on their coefficients. In order to be able to solve the equations with the standard PPN ansatz
\begin{equation}\label{eqn:standardppn}
\order{h}{4}_{00}=b_1\Phi_1+b_2\Phi_2+b_3\Phi_3+b_4\Phi_4+b_5\Phi_W+b_6 U^2+b_7\mathcal{A}+b_8\mathcal{B}\,,
\end{equation}
with constant coefficients $b_1$ to $b_8$, and the PPN potentials defined in~\cite{Will:1993ns}, the coefficients \(a_4, \ldots, a_{14}\) must be given by
\begin{gather}
a_4 = -4\pi b_4+8\pi b_8 \,,\quad
a_6 = -4\pi b_1-4\pi b_7 \,,\quad
a_7 = \frac{1}{4}b_2-\frac{1}{4}b_5+\frac{1}{2}b_6 \,,\quad
a_8 = -b_5 \,,\quad
a_9 = \frac{1}{2}b_2-\frac{5}{2}b_5+2b_6+\frac{1}{2}b_8 \,,\nonumber\\
a_5 = -4\pi b_3 \,,\quad
a_{10} = -2b_5 \,,\quad
a_{11} = \frac{1}{4}b_2-\frac{1}{4}b_5+\frac{1}{2}b_6+\frac{1}{2}b_8\,,\quad
a_{12} = -3b_5+b_6 \,,\quad
a_{13} = 0 \,,\quad
a_{14} = 8\pi b_7 \,,\label{eqn:standardcoeff}
\end{gather}
and are thus linearly dependent. The final gauge freedom in the resulting equations is resolved by choosing the standard PPN gauge \(b_8 = 0\).

We then must distinguish the two cases we found already at the second velocity order. For $C_3=C_4=0$, we find that \(\triangle\triangle\partial_i\order{\xi}{4}^i\) does not appear in the resulting field equations, and we can eliminate the spatial metric components by considering the linear combination
\begin{equation}
(2C_2 + C_5)\triangle\order{E}{4}^i{}_{i} + (C_2 - C_5)\triangle\order{E}{4}_{00} = 0
\end{equation}
However, we find that the equation depends on the undetermined term \(\order{\xi}{2}\), unless one imposes the additional condition
\begin{equation}
(C_2 + 2C_5)(C_1 + C_2 + 2C_5) =0\,.
\end{equation}
By doing so the coefficient in front of $\order{\xi}{2}$ vanishes and the fourth order equations can be solved independently of this tensor field. We find that the solution can be fully expressed in terms of the standard PPN ansatz~\eqref{eqn:standardppn} in this case.

In the generic case (i.e., $C_2C_3\neq C_4^2$) the second order tensor fields are defined unambiguously, and the fourth order field equations form a non-degenerate linear system of equations depending on the fields $\order{h}{4}_{00}$, $\order{h}{4}_{ij}$ and $\order{\xi}{4}^i$. To isolate the component \(\order{h}{4}_{00}\) we are interested in, we calculate the linear combination
\begin{equation}
\left(2C_2C_3-2C_4^2+C_3C_5\right)\triangle\order{E}{4}^i{}_{i}+\left(C_2C_3-C_4^2-C_3C_5\right)\partial_i\partial_j\order{E}{4}^{ij}+\left(-C_2C_3+C_4^2+C_3C_5+3C_4C_5\right)\triangle\order{E}{4}_{00}=0\,,
\end{equation}
which does not depend on \(\order{h}{4}_{ij}\) and \(\order{\xi}{4}^i\). The next step is to use the ansatz~\eqref{eqn:generppn} for the fourth order metric components and solve for the constant coefficients $a_4$ to $a_{14}$. To be able to calculate the PPN metric in the form~\eqref{eqn:standardppn} without defining generalized potentials, these coefficients must take the form~\eqref{eqn:standardcoeff}. This is the case if and only if
\begin{equation}
\left(C_1+C_2+C_3-2C_4+2C_5\right)\left(C_2C_3-C_4^2+2C_3C_5\right)\left(-C_2C_3+C_4^2+C_3C_5+3C_4C_5\right) = 0\,.
\end{equation}
If we don't impose one of these factors to vanish (e.g., for arbitrary $C_i$), we have to define additional potentials, which have already been used for the calculation of a superset of the PPN parameters in multimetric gravity~\cite{Hohmann:2013oca}. In this case we cannot express the fourth order metric perturbation in terms of the usual PPN parameters, which makes its interpretation more difficult, and requires further studies regarding the phenomenology of the new contributions.

\section{PPN parameters and classification of theories}\label{sec:classification}
Using the results from the previous section, we can now determine the PPN parameters as follows. First, using the second order metric perturbation~\eqref{eqn:metorder2}, one determines the PPN parameter \(\gamma\) as
\begin{equation}
\gamma = \frac{a_2}{a_1}\,.
\end{equation}
One then continues with the third order solution. To determine the PPN parameter $\alpha_1$, we use the gauge invariant formulation of the PPN formalism~\cite{Hohmann:2019qgo}. Here a different gauge is chosen, in which the third velocity order of the metric component $\order{h}{3}_{0i}$ only depends on the combination $V_i+W_i$ and reads
\begin{equation}
\order{h}{3}_{0i}=-\left(1+\gamma+\frac{\alpha_1}{4}\right)\left(V_i+W_i\right)\,.
\end{equation}
Therefore we can calculate the PPN parameter $\alpha_1$ by using the combination
\begin{equation}
\alpha_1 = -2(2 + 2\gamma + a_V + a_W)\,.
\end{equation}
Finally, we can use the fourth velocity order to determine the remaining PPN parameters from the metric ansatz~\eqref{eqn:standardppn}, as shown in full detail in~\cite{Will:1993ns}. Depending on the coefficients $C_i$ the PPN parameters are given in the following way.
\begin{enumerate}
\item
\framebox{\(C_5 = 0\)}: The field equations at the second order degenerate. Only one linear combination of the two coefficients in the second order metric components \(\order{h}{2}_{00}\) and \(\order{h}{2}_{ij}\) enters the field equations, so that one cannot solve for both components independently. The metric is not sufficiently determined by the field equations.
\item
\framebox{\(C_5 \neq 0\)}: The coefficients of the second order metric components \(\order{h}{2}_{00}\) and \(\order{h}{2}_{ij}\) enter the field equations independently.
\begin{enumerate}
\item
\framebox{\(C_2C_3 = C_4^2\)}: The field equations at the second order degenerate. \(\order{\xi}{2}^i\) is not uniquely determined by the field equations.
\begin{enumerate}
\item
\framebox{\(C_3 \neq 0\)}: The metric components \(\order{h}{2}_{00}\) and \(\order{h}{2}_{ij}\) depend on the undetermined tensor field \(\order{\xi}{2}^i\). The system cannot be solved uniquely for these metric components.
\item
\framebox{\(C_3 = 0\)}: The metric components \(\order{h}{2}_{00}\) and \(\order{h}{2}_{ij}\) are independent of the undetermined tensor field \(\order{\xi}{2}^i\) and can be solved for seperately. The PPN parameter \(\gamma\) determined from this solution reads
\begin{equation}
\gamma = \frac{C_2 - C_5}{2C_2 + C_5}\,.
\end{equation}
At the third order, one obtains \(\alpha_1 = 0\).
\begin{enumerate}
\item
\framebox{\(C_2 + 2C_5 = 0\)}: The fourth order can be solved independently of \(\order{\xi}{2}^i\). The PPN parameters are
\begin{equation}
\beta = \gamma = 1\,.
\end{equation}
\item
\framebox{\(C_1 + C_2 + 2C_5 = 0\)}: The fourth order can be solved independently of \(\order{\xi}{2}^i\). The PPN parameters are
\begin{equation}\label{eqn:ppnbad1}
\beta = \frac{7C_1 + 12C_5}{8C_1 + 12C_5} = 1 - \frac{C_1}{8C_1 + 12C_5}\,, \quad
\gamma = \frac{C_1 + 3C_5}{2C_1 + 3C_5} = 1 - \frac{C_1}{2C_1 + 3C_5}\,.
\end{equation}
Note that \(C_1 \neq 0\), and hence \(\beta \neq 1\) and \(\gamma \neq 1\), unless also the previous condition is satisfied.
\item
\framebox{otherwise}: The fourth order cannot be solved independently of the undetermined component \(\order{\xi}{2}^i\).
\end{enumerate}
\end{enumerate}
\item
\framebox{\(C_2C_3 \neq C_4^2\)}: The field equations at the second order form a non-degenerate linear system and can be solved independently for \(\order{h}{2}_{00}\), \(\order{h}{2}_{ij}\) and \(\order{\xi}{2}^i\). The PPN parameter \(\gamma\) determined from this solution reads
\begin{equation}\label{eqn:ppnbadg3}
\gamma = \frac{C_3(C_2 - C_5) - C_4^2}{C_3(2C_2 + C_5) - 2C_4^2} = 1 - \frac{C_3(C_2 + 2C_5) - C_4^2}{C_3(2C_2 + C_5) - 2C_4^2}\,.
\end{equation}
At the third order, one obtains \(\alpha_1 = 0\).
\begin{enumerate}
\item
\framebox{\(C_3(C_2 + 2C_5) = C_4^2\)}: The field equations at the fourth order yield the PPN parameters
\begin{equation}
\beta = \gamma = 1\,.
\end{equation}
\item
\framebox{\(C_2C_3 = C_4^2 + C_3C_5 + 3C_4C_5\)}: The field equations at the fourth order yield the PPN parameters
\begin{equation}\label{eqn:ppnbad2}
\beta = \frac{3C_3 + 7C_4}{4C_3 + 8C_4} = 1 - \frac{C_3 + C_4}{4C_3 + 8C_4}\,, \quad
\gamma = \frac{C_4}{C_3 + 2C_4} = 1 - \frac{C_3 + C_4}{C_3 + 2C_4}\,.
\end{equation}
It further follows that \(\beta \neq 1\) and \(\gamma \neq 1\), since \(C_3 + C_4 \neq 0\), unless also the previous condition is satisfied.
\item
\framebox{\(C_1 + C_2 + C_3 - 2C_4 + 2C_5 = 0\)}: The field equations at the fourth order yield the remaining PPN parameter
\begin{equation}\label{eqn:ppnbadb3}
\beta = \frac{7C_2C_3 - 7C_4^2 + 2C_3C_5}{8C_2C_3 - 8C_4^2 + 4C_3C_5} = 1 - \frac{C_2C_3 - C_4^2 + 2C_3C_5}{8C_2C_3 - 8C_4^2 + 4C_3C_5}\,.
\end{equation}
Also in this case \(\beta \neq 1\) and \(\gamma \neq 1\) unless a previous condition is satisfied.
\item
\framebox{otherwise}: The field equations at the fourth order cannot be solved by the standard PPN metric.
\end{enumerate}
\end{enumerate}
\end{enumerate}

\begin{figure}[htbp]
\begin{tikzpicture}[decision/.style={draw,rectangle,fill=black!10!white,outer sep=2pt},result/.style={draw,rectangle,rounded corners,outer sep=2pt},every edge/.style={draw,-Stealth},good/.style={fill=green!25!white},maybe/.style={fill=yellow!25!white},noppn/.style={fill=red!25!white},degen/.style={fill=magenta!25!white}]
\node[decision] (c0) at (0,0) {$C_0$};
\node[decision] (c5) at (0,-2) {$C_5$};
\node[decision] (c2c3c4) at (0,-4) {$C_2C_3 - C_4^2$};
\node[decision] (c3) at (-4,-4) {$C_3$};
\node[decision] (c2c5) at (-4,-6) {$C_2 + 2C_5$};
\node[decision] (c1c2c5) at (-4,-8) {$C_1 + C_2 + 2C_5$};
\node[decision] (c3c2c5c4) at (5,-4) {$C_2C_3 + 2C_3C_5 - C_4^2$};
\node[decision] (c3c2c4c5c4) at (5,-6) {$C_2C_3 - C_3C_5 - 3C_4C_5 - C_4^2$};
\node[decision] (c1c2c3c4c5) at (5,-8) {$C_1 + C_2 + C_3 - 2C_4 + 2C_5$};
\node[result,noppn] (nominkbg) at (4,0) {no Minkowski background};
\node[result,degen] (h2undet) at (-4,-2) {$\order{h}{2}$ undetermined};
\node[result,good] (ppngood) at (0,-6) {$\beta = \gamma = 1$};
\node[result,maybe] (ppnbad) at (0,-8) {$\beta \neq 1, \gamma \neq 1$};
\node[result,degen] (h4undet) at (-4,-10) {$\order{h}{4}$ undetermined};
\node[result,noppn] (noppn) at (5,-10) {beyond PPN solution};
\path (0,1) edge[ultra thick] (c0);
\path (c0) edge node[above, near start] {$\neq 0$} (nominkbg);
\path (c0) edge[ultra thick] node[left, near start] {$= 0$} (c5);
\path (c5) edge node[above, near start] {$= 0$} (h2undet);
\path (c5) edge[ultra thick] node[left, near start] {$\neq 0$} (c2c3c4);
\path (c2c3c4) edge[ultra thick] node[above, near start] {$= 0$} (c3);
\path (c2c3c4) edge node[above, near start] {$\neq 0$} (c3c2c5c4);
\path (c3) edge node[left, near start] {$\neq 0$} (h2undet);
\path (c3) edge[ultra thick] node[left, near start] {$= 0$} (c2c5);
\path (c2c5) edge[ultra thick] node[above, near start] {$= 0$} (ppngood);
\path (c2c5) edge node[left, near start] {$\neq 0$} (c1c2c5);
\path (c1c2c5) edge node[above, near start] {$= 0$} (ppnbad);
\path (c1c2c5) edge node[left, near start] {$\neq 0$} (h4undet);
\path (c3c2c5c4) edge node[above left, near start] {$= 0$} (ppngood);
\path (c3c2c5c4) edge node[right, near start] {$\neq 0$} (c3c2c4c5c4);
\path (c3c2c4c5c4) edge node[above left, near start] {$= 0$} (ppnbad);
\path (c3c2c4c5c4) edge node[right, near start] {$\neq 0$} (c1c2c3c4c5);
\path (c1c2c3c4c5) edge node[above, near start] {$= 0$} (ppnbad);
\path (c1c2c3c4c5) edge node[right, near start] {$\neq 0$} (noppn);
\end{tikzpicture}
\caption{Full classification of generalized newer general relativity theories. The path highlighted by thick arrows corresponds to STEGR. Theories with \(\beta = \gamma = 1\) are in full agreement with observations. Theories with deviating PPN parameters receive bounds on their parameters, and are still in agreement if these bounds are met. Other classes of theories are either degenerate, so that not all components of the metric perturbation are determined by the perturbative field equations, or cannot be solved within the standard PPN formalism.}
\label{fig:class}
\end{figure}
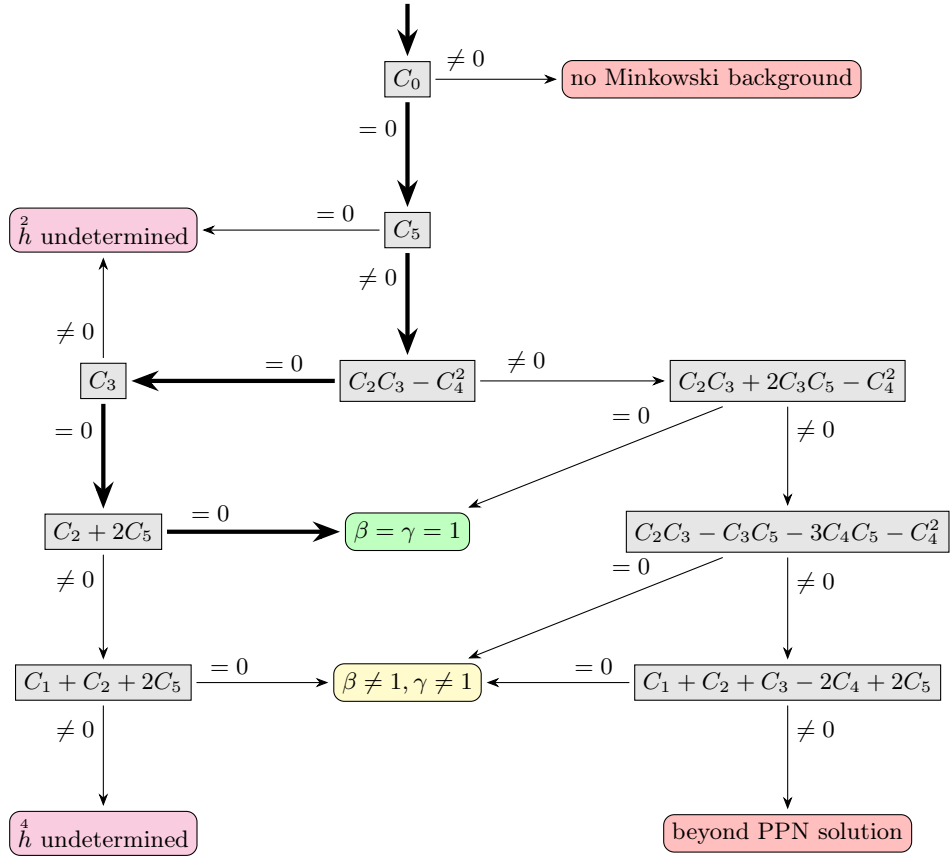

In all cases where the full set of PPN parameters can be found, the theory is fully conservative, i.e., only \(\beta\) and \(\gamma\) potentially deviate from the general relativity values \(\beta = \gamma = 1\), while all other PPN parameters vanish,
\begin{equation}
\alpha_1 = \alpha_2 = \alpha_3 = \zeta_1 = \zeta_2 = \zeta_3 = \zeta_4 = \xi = 0\,.
\end{equation}
The complete classification is shown in figure~\ref{fig:class}. In summary, we find two families of theories whose PPN parameters agree with those of general relativity: a two-parameter family, satisfying
\begin{equation}\label{eqn:degengood}
C_3 = C_4 = C_2 + 2C_5 = 0\,, \quad C_5 \neq 0\,,
\end{equation}
parametrized by \(C_1\) and \(C_5\), and a four-parameter family
\begin{equation}
C_2C_3 + 2C_3C_5 - C_4^2 = 0\,, \quad C_5 \neq 0\,, \quad C_2C_3 - C_4^2 \neq 0\,,
\end{equation}
parametrized by \(C_1, C_2, C_3, C_5\). Note in particular that STEGR, as well as the \(f(Q)\) class of theories, satisfy the condition~\eqref{eqn:degengood}, and thus fall into this class. Further, we found three families of theories whose PPN parameters~\eqref{eqn:ppnbad1}, \eqref{eqn:ppnbad2} as well as~\eqref{eqn:ppnbadg3} and~\eqref{eqn:ppnbadb3} generically deviate from the general relativity values, but can be brought arbitrarily close to these values by choosing the parameters \(C_1, \ldots, C_5\) such that they become sufficiently close to the values for the two aforementioned families. Both classes of theories therefore merit further studies.

\section{Conclusion}\label{sec:conclusion}
We have calculated the parametrized post-Newtonian limit of a general class of symmetric teleparallel gravity theories, whose Lagrangian is a free function of the five parity-even scalar invariants which are quadratic in the non-metricity tensor. As a result, we have obtained a full classification of these theories into several classes, depending on the Taylor coefficients of the Lagrangian function. Most notably, we found two families of theories whose PPN parameters are identical to those of general relativity, such that these theories are indistinguishable from the latter by measurements of the PPN parameters. Further, we found several classes of fully conservative theories, which means that only the PPN parameters \(\beta\) and \(\gamma\) deviate from their general relativity values. For these classes, the general relativity values can be approximated to arbitrary precision by an appropriate choice of the Lagrangian function, such that certain theories within these classes are still compatible with observations. Finally, we found other classes of theories, whose PPN limit is either undetermined by the perturbative field equations or does not possess the form assumed by the PPN formalism.

Our work lays the foundation for applying the PPN formalism to symmetric teleparallel gravity theories. Using the post-Newtonian expansion of the non-metricity tensor which we presented here, one may now calculate the PPN parameters for other theories, such as scalar-non-metricity theories~\cite{Jarv:2018bgs,Runkla:2018xrv} and generalizations. For torsional teleparallel theories, an analogous step has been undertaken in~\cite{Ualikhanova:2019ygl} and~\cite{Emtsova:2019qsl,Flathmann:2019khc}. Further, one may consider more general teleparallel gravity theories, which feature both torsion and non-metricity~\cite{Jimenez:2019ghw}, or which include derivative couplings between a scalar field and non-metricity, in analogy of a torsional extension to Horndeski gravity~\cite{Bahamonde:2019shr,Bahamonde:2020cfv}.

Another line of further research is by studying other physical aspects of symmetric teleparallel theories of gravity, such as the gravitational waves emitted from compact binary systems~\cite{Blanchet:2013haa} or perturbed compact objects~\cite{Kokkotas:1999bd,Berti:2009kk}, as well as cosmological perturbations~\cite{Mukhanov:1990me}. Correlating these results obtained from these further studies with other properties, such as gravitational wave propagation~\cite{Hohmann:2018xnb,Hohmann:2018wxu} or a Hamiltonian analysis~\cite{DAmbrosio:2020njs}, would provide an effective tool to constrain the large class of symmetric teleparallel gravity theories.

\begin{acknowledgments}
KF gratefully acknowledges support by the DFG within the Research Training Group \textit{Models of Gravity}. MH gratefully acknowledges the full financial support by the Estonian Research Council through the Personal Research Funding project PRG356 and by the European Regional Development Fund through the Center of Excellence TK133 ``The Dark Side of the Universe''. This article is based upon work from COST Actions CANTATA (CA15117) and QGMM (CA18108), supported by COST (European Cooperation in Science and Technology).
\end{acknowledgments}

\bibliography{teleppn}

\end{document}